# Doubly Triggered Conductance Across Thin Zinc Oxysulfide Films*

A. Givon,[1,a)] K.X. Steirer,[2] E. Segre[3] and H. Cohen[1,*)]

[1]*Department of Chemical Research Support, Weizmann Institute, Rehovot, 7610001, Israel*

[2]*Department of Physics, Colorado School of Mines, Golden, CO, 80214, USA*

[3]*Faculty of Physics, Weizmann Institute, Rehovot, 7610001, Israel*

Chemically resolved electrical measurements (CREM) of zinc oxysulfide (ZnOS) over-layers on gold show very poor conductance under either electrical or optical input signals, whereas simultaneous application of the two yields extremely high sample currents. The effect and its dependence on wavelength and electrical parameters is explained by the in-situ derived band diagram, in which a buffer level of charge traps cannot contribute directly to conductance, however amplifies the photoconductance by orders of magnitudes under sub-bandgap illumination. This AND-type doubly-triggered response proposes interesting applications and an answer to problems encountered in related optoelectronic devices.

Thin film solar cells, such as copper indium diselenide (CIS) and copper zinc tin sulfide (CZTS), exploit buffer layer materials to conduct charges from the photoactive domain to the transparent conducting oxide. ZnOS is a potential candidate thanks to its wide band-gap, scalable and facile deposition and typically low waste and safety costs.[1] It further shows promise in nanorod based self-powered UV detectors and photocatalytic bacterial inactivation.[2–4] However, application of ZnOS in photovoltaics and light emitting devices is challenged by high recombination rates and low extrinsic doping efficiency.[5]

The understanding of ZnOS structure and properties still bears open questions. The ZnOS film conductivity is highly sensitive to subtle changes in the chemical bath deposition (CBD), while typically, these films are insulating.[6] For earth abundant photovoltaics, such as CZTS and CIS, recent reports have noted a too large conduction band offset between the absorber layer and the ZnOS buffer, combined with significant photocurrent, which indicates the likelihood of an alternate conduction path for photogenerated charges.[7,8] The ZnOS photoluminescence (PL) consists of an intense line around 2.5 eV and, yet, in spite of extensive related study, its assignments are debatable.[9–11] Deeper understanding of the optoelectronic properties of CBD ZnOS is therefore a challenge and a key towards improved processing and performance.

Based on X-ray photoelectron spectroscopy (XPS), it is often possible to resolve in a noncontact manner the electrostatic potential at selected domains. As previously demonstrated by chemically resolved electrical measurements (CREM), the electron analyzer acts then as a chemically selective voltmeter,[12–15] allowing for the subtleties of charge trapping and internal fields in heterojunctions to be identified.[16] Here, we exploit recent instrumental developments in our CREM setup for the study of ZnOS. The band diagram is constructed in-situ and leveraged to probe both electrical and optical properties of the system. We thus map occupied and unoccupied energy levels in the band diagram and, notably, reveal an unusually high, doubly-triggered conductivity, the triggers of which being light illumination and hot-electron injection. These findings help to explain the ZnOS operation in solar cell interfaces and raise new opportunities for defect enhanced UV sensing.

The thin film ZnOS CBD process is detailed in an earlier report.[7] For the photoluminescence (PL), bandgap excitation was supplied with a HeCd laser and 325 nm razor edge long pass filter at 5 mW. The spot size was 250 μm. Detection by an Oriel 78235 with the 300/500 grating at 500 nm center wavelength and 280 μm slit was taken in one second exposure times. PL data was corrected for the CCD quantum efficiency and smoothed using a cubic spline procedure. All measurements were made at room temperature.

XPS and CREM were performed on a slightly modified Kratos Ultra-DLD spectrometer (see inset to Fig. 1a), using a monochromatic Al Kα source at low power, 15-75 W. The gold substrate of the sample was grounded with a double-side conductive carbon tape. The electron flood gun (eFG) was operated at 2-5 V bias and filament current of 1.8-2A. Sample bias was varied in the range of 0-5 V. The kinetic energy of incoming electrons refers to the conditions at which the local vacuum level at the sample's surface and the eFG coincide.[17] The eFG has no direct line of sight between the filament and the sample. Hence, sample's response to external light sources is decoupled from light shined by the eFG. Light was supplied by means of a Prismatix beam combiner, connected to a triple source with 630 nm, 365 nm and white LEDs. For each selected bias and LED wavelength, an Off-On-Off illumination sequence was performed, such that any irreversibility was documented. Repeated measurements were conducted at selected surface charge conditions, in order to reveal sample's instabilities.

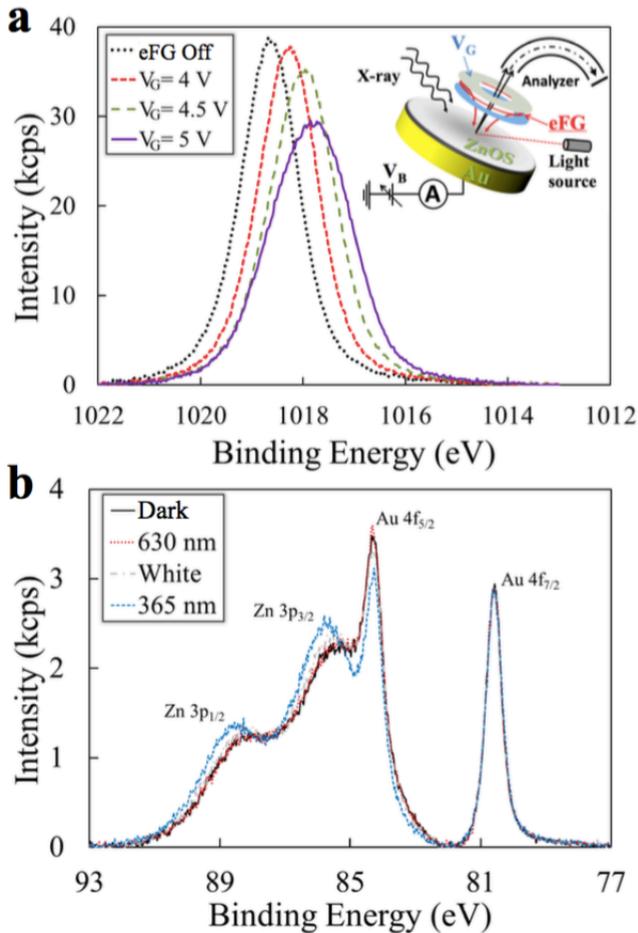

FIG. 1. (a) The Zn $2p_{3/2}$ spectrum, demonstrating peak shifting under controlled eFG conditions. The inset illustrates the CREM setup, where input signals include the x-ray source, the eFG, the light source and the bias on sample back contact. Output signals are the sample current, measured on the back contact, and the photoelectron spectrum measured at the exit of the analyzer. (b) The Au 4f and Zn 3p spectral window at eFG on and different light illumination conditions, showing that the grounded Au substrate retains stable electrostatic potential while the ZnOS layer charges up.

Fig. 1a exemplifies CREM core-level shifts of the Zn $2p_{3/2}$ line, recorded under sample bias of -3.5 V and selected eFG settings ($V_G$ = off, 4, 4.5 and 5V). These line-shifts directly reflect the changes in surface potential when exposed to a flux of external electrons. In addition to their peak shift, the spectral lines in Fig. 1a undergo broadening, a feature typical to insulating layers across which potential gradients emerge under the external stimuli.[18] Fig. 1b presents both the Au 4f and Zn 3p doublets, as recorded under different light illumination settings (dark, red, white and UV), while the eFG is kept at fixed conditions (with $V_G$=5 V). The substrate signal (Au in Fig. 1b) does not shift at all, because no potential changes evolve in the grounded metal. The O 1s, C 1s and S 2p (not shown) yielded line shifts that were generally consistent with the Zn-derived data.

In the following, we denote $\Delta_L$ for changes detected upon switching on the light source (under fixed, pre-determined eFG conditions) and $\Delta_e$ for changes detected under switching on the eFG (at fixed, pre-determined, light conditions). The sign of the potential is determined such as to follow the changes in surface energy: positive $\Delta V$ values correspond to increased negative charge.

A set of CREM data is summarized in Fig. 2, showing changes in the sample current ($\Delta I$) and surface potential ($\Delta V$), as measured under selected eFG and illumination conditions. Fig. 2a presents the photo response ($\Delta_L V$ and $\Delta_L I$) under 'eFG-off' conditions (no electron flux in), demonstrating overall extremely low values. Note the slight yet significant differences in $\Delta_L I$ and elemental $\Delta_L V$ of the three light sources: Higher values are obtained for the white (W) source, as compared to the red (and UV) source. These differences hint on the presence of states within the wide bandgap, $E_g$ = 3.8 eV, as further discussed below.

Complementary to the photoresponse in Fig. 2a, Fig. 2b demonstrates the sample's response to the eFG ($\Delta_e V$ and $\Delta_e I$) under dark. High eFG-induced potentials (data extracted from the Zn $2p_{3/2}$ line) and low sample currents are observed. Both features are typical to insulating media, however not at all trivial in CREM, where the injected electrons are of relatively elevated energy, above the vacuum level. Therefore, the electron probability to traverse a thin (10-15 nm) layer via its conduction band is typically high, even for insulating layers.[19] Here, the sample exhibits up to ~30 nA under eFG conditions that can deliver ~2 μA, which suggests efficient charge trapping: The accumulation of charge would instantly raise the surface energy, such as to repel the incoming electrons and, eventually, stabilize steady-state conditions for which the current through the sample is very low and the surface potential is close to the one set at the eFG, as indeed observed here.

The response to two input signals, both eFG and light illumination, can be followed in Figs. 2c and 2d, where $\Delta_e V$ and $\Delta_e I$ are depicted for each of the light sources under selected (fixed) eFG conditions: eFG off and six 'on' settings: $V_G$= 2.2, 3.0, 3.5, 4.0, 4.5 and 5.0 V. A curve is plotted also for the final (repeated) measurement at dark. All curves in Fig. 2c obey a roughly linear dependence, $\Delta_e I$ vs $I_{dark}$. Also (not shown) $\Delta_L I$ and $\Delta_L V$ are linearly correlated, which is typical to photo-induced conductivity. Remarkably, current magnitudes much higher than those in Fig 2a and 2b are observed. In addition, the differences between curves are pronounced, which suggests selective and controllable activation of transport channels.

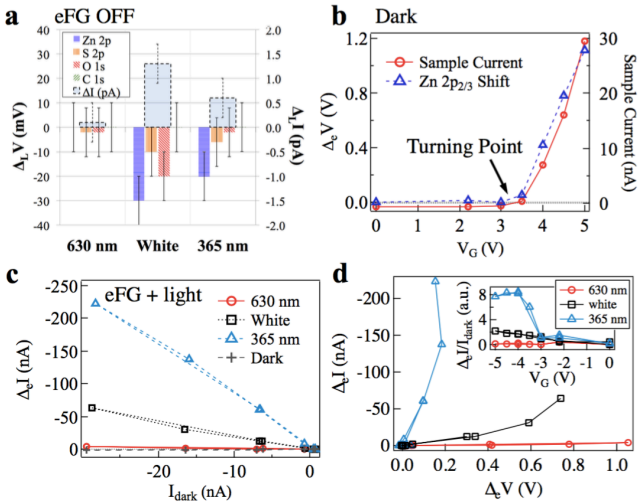

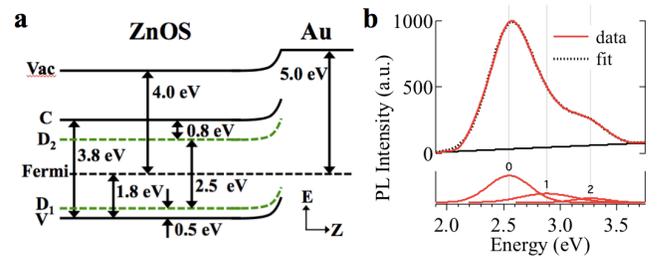

FIG. 2. CREM-derived current and potential changes under: light illumination with eFG off (a); eFG application at dark, as a function of eFG bias, $V_G$ (b); applying both light and eFG input signals (c) and (d), where (c) demonstrates the response to eFG under given illumination conditions (note the nearly linear $\Delta_e I$ vs I functionality) and (d) presents both current and potential data. The inset to (d) presents normalized photocurrents as a function of eFG bias.

FIG. 3. (a) The ground state band diagram of the system, including defect levels and interface band bending. (b) The PL spectrum of ZnOS thin film and the three components obtained from its curve fitting (using Voigt line shapes).

It should be noted that several data points in each of the curves in Fig. 2c gather near the origin, where $V_G$ is below a critical value (the turning point) and landing of eFG electrons on the surface is disabled. A useful zoom into the region below and around the turning point is presented in Fig. 2d inset. It depicts normalized current values, $\Delta_e I/I$, as a function of $V_G$. A step-like shape characterizes the UV curve: below the $V_G$ turning point, it is practically zero and above that point it retains a roughly constant value. This is a clear demonstration of a transport channel that requires two triggering input signals to switch on: electron injection and UV illumination. In contrast, the W source undergoes a monotonic increase across the turning point and the R curve retains a nearly constant (non-vanishing) value.

The results in Fig. 2 can be explained by means of the band diagram shown in Fig. 3a. We directly extract the valence band edge and the work function (WF) at charging-free conditions, as described elsewhere.[16,18] These values indicate that the system's Fermi level is slightly below mid-gap, accompanied by large band-bending near the ZnOS/Au interface, ca. 1 eV in magnitude. The width of the related space charge is not known to us, but as shown below, our data suggest a rather narrow region, on the scale of 1 nm. In addition, an optically determined band gap energy of 3.8 eV is used for the band diagram,[7] with standard deviation of ±0.17 eV.[7] Complementary information is provided by our ex-situ PL measurements; see Figure 3b, where a dominant PL peak at ~2.55 eV and two additional peaks at 2.9 and 3.3 eV are resolved.

Consistent with the entire set of CREM and PL data, the band diagram of ZnOS/Au includes two dominant trap states, $D_1$ and $D_2$, located respectively at ~0.5 eV above the top of valence band and ~0.8 eV below the bottom of conduction band. Both levels are deep as compared to room temperature, and $D_1$ appears to be empty of electrons, except for a space charge region near the gold interface, across which band bending evolves. Both $D_1$ and $D_2$ are capable of electron trapping, but the lifetime of trapped electrons in $D_2$ is by far shorter than in $D_1$, as verified from the PL data and the CREM results in Fig. 2. Importantly, even when partially re-filled by electrons, $D_1$ can hardly discharge them to ground, because of the ~1 eV interface barrier. Transport of these electrons to ground is realized here only via photoexcitation to the $D_2$ states and the conduction band (C). Note that the very low eFG-induced currents measured at dark, Fig. 2b, yield strong evidence for rapid and efficient trapping of its (hot) electrons by the $D_1$ states. In other words, $D_1$ acts as an efficient capacitive component from which the trapped charge can be optically fed into higher levels.

At least two independent results suggest that $D_1$ and $D_2$ are spatially very close. First, the PL intensity of this transition is by large higher than the competing $D_2$-V and C-$D_1$ transitions, in spite of the fact that extended band states should enable spatial overlap with any trap level. Second, a broad tail towards low energies in the PL spectrum is seen in Fig. 3b, down to 2.1 eV, which is typical of spatially confined excitonic states. Consequently, high efficiency of optical $D_1$ to $D_2$ pumping is expected. Rough estimates point to rather high concentrations of the $D_1$ states, above $10^{19}$ cm$^{-3}$.

The interface band bending indicated in Fig. 3a is about 1 eV in height. As already stated, it plays an important role against transport of electrons to the substrate. Hence, any $D_1$-trapped electrons are restricted to remain in the layer. We further propose that the interface band bending itself consists of electron-filled $D_1$ states. On the other hand, electrons in $D_2$ or C are insensitive to that barrier. Hence, electrons injection to $D_1$ would open new channels for photo-activity. A summary of the photon energies used here and of transitions predicted by the band diagram is given in Fig. 4a. Clearly, with no electrons captured in $D_1$, our red source does not match any allowed transitions. The white source, however, is just slightly below resonance with the V-$D_2$ transition. Therefore, even under eFG off conditions, a weak yet significant photo response to white light can be observed, Fig. 2a.

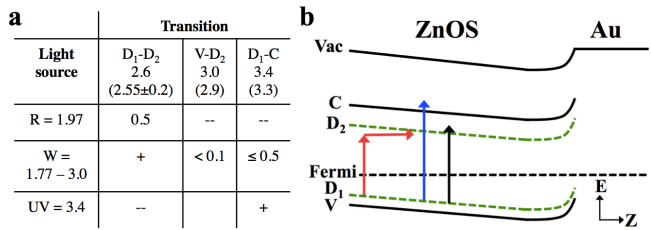

FIG. 4. (a) Table of compatibility between expected electron transitions as predicted by the band diagram in figure 3a and the photon energies (in eV) supplied by our light source. (b) The distorted band diagram, as realized under eFG-induced surface charging. Light excitations from $D_1$ are indicated, including illustration of a transition indirect in space.

Remarkably, with the combined inputs, eFG plus light, transitions from $D_1$ are switched on by the emerging occupation of electrons in $D_1$. Based on Fig. 4a, one expects the UV source to activate resonant $D_1$-C transitions, as indeed observed experimentally. However, the red source is slightly below the $D_1$-$D_2$ transition and the white source, which covers the $D_1$-$D_2$ transition, cannot apply for $D_1$-C excitations. In this matter, one should recall smearing effects like the energy width of levels and, importantly, the fact that the band diagram is distorted under charge trapping, such that in-layer fields emerge upon filling up $D_1$. As illustrated in Fig. 4b, in-layer fields can enable spatially indirect transitions with sub-resonance photon energies, provided that the excitation involves $D_1$ and $D_2$ traps located along the field lines. Thus, the normalized current under red light illumination, inset to Fig. 2d, manifests a small yet consistent deviation from a constant value, which reflects the rise in spatially eligible trap sites for $D_1$-$D_2$ transitions. A similar tendency is realized under white light illumination but with a larger slope, because the effect applies for two channels: V-$D_2$ and $D_1$-$D_2$. In fact, a close look into the data suggests, as predicted by the band diagram, that the slope for white light increases as soon as $D_1$ states start to fill up with electrons, above the turning point in Fig. 3d inset. Finally, for the UV light, the step-like change seen in Fig. 2d inset aligns with the threshold condition for eFG electrons to reach the sample. Then, as the $D_1$-C channel opens up, the photocurrent simply scales with the amount of trapped (in $D_1$) charge. Slight deviations from a perfectly constant value are due to the increased kinetic energy of incoming electrons when $V_G$ is elevated.

Various applications of the double triggering may be thought of. An attractive example regards self powered photodetectors which rely on small changes in Schottky barriers for high photosensitivity.[2] The defect coupled conductivity mechanism revealed here by CREM provides an alternative explanation for the performance enhancement observed there.

As a final remark, the methodology used here, involving two input signals, may be compared with two-photon techniques, in which intermediate excited states are exploited as well. There are yet several important differences realized by the CREM approach. First, the bandgap states are scanned here by combined optical and electrical means, such that field is built controllably within the film, introducing complementary features not accessible by the two-photon techniques. Second, in addition to sample's current, CREM directly probes the actual potentials developing at the layer, including inner profile details, such that the driving *transport* mechanisms can be followed directly. Third, the in-situ complementary data for band diagrams proposes important advantages for any charge occupation and transport analysis. Having in-situ chemical analysis and no top contact issues is also a clear advantage, allowing real-time follow-up of material instabilities. In this respect, CREM offers key advantages over standard electrical measurements as well.[20]

In summary, the double triggering mechanism revealed here proposes useful applications for ZnOS, not just as an AND-type switching means, but also for fine control over layer conductance: The $D_1$ level provides a tunable magnifier that does not contribute directly to the conductance (and, in fact, even blocks the transport of hot electrons), however with a second input signal, it provides the control over output of other channels. This control is twofold in essence, implying level-occupation and emerging potential gradients. Thus, the combined eFG plus light illumination tool applies for both the activation and characterization of the system.

We thank Pat Dippo for performing the PL measurements and the Israel Science Foundation for financial support, project 1522. KXS acknowledges the Director's Postdoctoral Fellowship program at the National Renewable Energy Lab for support.